\documentclass[a4paper,11pt]{article}
\usepackage{jheppub}
\usepackage{eurosym}
\usepackage{amssymb}
\usepackage{amsfonts,cite}
\usepackage{amsmath}
\usepackage{graphicx}
\usepackage{multirow}
\usepackage{epsfig}
\usepackage[utf8]{inputenc}
\usepackage{multicol}
\usepackage{graphicx}
\usepackage{fancyhdr}
\usepackage{wrapfig}
\usepackage{placeins}
\usepackage{slashed}
\usepackage{lipsum}
\usepackage{array}
\usepackage{amscd}
\usepackage{tikz-cd}
\usepackage[thinlines]{easytable}
\usepackage{mathrsfs}
\usepackage[export]{adjustbox}
\usepackage{blindtext}

\newbox\mybox

\newcommand\fverb{\setbox\mybox=\hbox\bgroup\verb}
\newcommand\fverbdo{\egroup\medskip\noindent\fbox{\unhbox\mybox}\ }
\newcommand\fverbit{\egroup\item[\fbox{\unhbox\mybox}]}

\abstract{We study a time-dependent non-Hermitian generalisation of the Sch\"utte-Da~Provid\^encia  model describing a bosonic mode coupled to collective particle-hole excitations. Using a time-dependent Dyson map, we construct a Hermitian counterpart and reduce the collective fermionic sector by means of the random phase approximation (RPA). The resulting dynamics is mapped to two time-dependent harmonic-oscillator branches with instantaneous RPA frequencies $W_\pm(t)$. We determine the corresponding stability regions and compute transition probabilities between instantaneous oscillator states. In first-order instantaneous-basis perturbation theory the leading transition $n\to n+2$ is proportional to $\dot W_j/W_j$, showing that it is purely nonadiabatic and absent in the time-independent case. We compare this result with exact Lewis-Riesenfeld transition amplitudes within the RPA approximation. Numerical examples show that different components of the Dyson map provide distinct driving mechanisms: the scaling parameter modulates the effective coupling, while the squeezing parameter acts through a moving-boundary contribution. In both cases the induced collective transitions exhibit parametric-resonance peaks and sideband structures.}


\title{Driving collective RPA modes by a time-dependent Dyson map}

\author[a]{Andreas Fring}
\author[b]{Marta Reboiro}

\affiliation[a]{Department of Mathematics, City St George's, University of London,  Northampton Square, \\ London EC1V 0HB, UK}
\affiliation[b]{Institute of Physics of La Plata (IFLP), Boulevard 113 \& 63,  La Plata C.P. 1900, Argentina}

\emailAdd{a.fring@city.ac.uk}	
\emailAdd{reboiro@fisica.unlp.edu.ar}


\begin{document}
	\maketitle
	
	\pagestyle{fancy}
	\fancyhead{} 
	\fancyhead[LE,RO]{\small\itshape  Driving collective RPA modes by a time-dependent Dyson map} 
	
	\renewcommand{\headrulewidth}{0.4pt}

\section{Introduction}

In a recent work \cite{fring2026induced} we applied the general framework devised for the effective description of non-Hermitian Hamiltonian systems \cite{Bender:1998ke,Benderrev,AliI,Alirev} to a time-dependent non-Hermitian
Sch\"utte-Da~Provid\^encia spin-boson model \cite{schutte1977solvable} with moving-boundary interpretation. For explicitly time-dependent systems this structure becomes more subtle, since the
time-dependence of the Dyson map contributes an additional gauge-like term to the
Hermitian Hamiltonian \cite{fringmoussa,fring2023introPTt}. Time-independent pseudo-Hermitian versions of this model have been considered previously in \cite{reboiro2022qu}. For the setting considered in \cite{fring2026induced} the time-dependent Dyson map contained a bosonic squeezing transformation, whose time derivative produced a dilatation term. This term has the same structure as the inertial
contribution that appears when a Hermitian system on a moving interval is transformed to
a fixed reference interval. The main consequence was dynamical rather than spectral. In
the fixed-boundary Hermitian model the operator $Q=N-S_0$ is conserved, so that the
spin-boson Hilbert space decomposes into independent sectors and transitions changing
the boson number by two are forbidden. The moving-boundary term breaks this
conservation law and opens $n\leftrightarrow n\pm2$ channels. For closed boundary
protocols with constant background parameters the first-order integrated amplitude
vanishes, whereas a time-dependent non-Hermitian deformation can change the dressed
basis during the motion and thereby suppress or enhance the accumulated transition
amplitude by coherent interference.

The present paper is a continuation of that analysis, but with a different emphasis. We
do not study transitions between the finite-dimensional spin-boson sectors of the
spin-$1/2$ model. Instead, we consider the collective many-fermion version of the
Sch\"utte-Da~Provid\^encia model and reduce it by means of the random phase
approximation (RPA) \cite{bohm1951collective,pines1952collective,bohm1953collective,Neer16,Berkelbach,Hellgren,Folprecht}, see e.g. \cite{co2023introducing} for a recent review. The relevant degrees of freedom are then not individual spin-boson doublets, but collective particle-hole excitations of two degenerate fermionic shells coupled
to a bosonic mode. After the Dyson transformation and the RPA reduction, the problem
is mapped to time-dependent harmonic-oscillator branches. The central question therefore
becomes how a time-dependent Dyson map can drive transitions between collective RPA
oscillator states.

Originally the standard Sch\"utte-Da~Provid\^encia model was introduced as a solvable model of
boson condensation. It describes a bosonic mode coupled to collective particle-hole
excitations between two degenerate fermionic shells. The fermionic degrees of freedom can
be represented by collective $\mathfrak{su}(2)$ quasi-spin generators, with $S_+$ creating a
particle-hole excitation and $S_-$ annihilating one. This makes the model a useful
many-body analogue of spin-boson and light-matter Hamiltonians, while retaining enough
algebraic structure to allow explicit calculations. In the collective regime, where the
occupation of particle-hole excitations is small compared with the shell degeneracy, the
Holstein-Primakoff representation  \cite{Holstein} provides a natural route to a bosonic description of the
fermionic sector.

The RPA is precisely designed for such small-amplitude
collective dynamics. In its simplest form, it linearises the collective particle-hole motion
around a reference configuration and replaces the relevant collective excitations by bosonic
normal modes. It is widely used in many-body physics to describe collective vibrations,
response functions and stable small oscillations of interacting systems \cite{rowe1968equations,rowe2010nuclear,mahan2013many,blaizot1986quantum,co2023introducing}. In the present
setting the RPA approximation converts the quasi-spin sector into an effective bosonic
degree of freedom. The Dyson-transformed Sch\"utte-Da~Provid\^encia Hamiltonian then
reduces to a quadratic bosonic Hamiltonian, whose normal modes are determined by an
instantaneous RPA eigenvalue problem.

Introducing time-dependence and non-Hermiticity is motivated by two related points.
First, time-dependent non-Hermitian Hamiltonians naturally arise as effective generators
for driven systems, systems with controlled gain and loss, and systems described in
time-dependent non-orthogonal representations. Secondly, in the quasi-Hermitian setting
the time-dependence of the Dyson map is not merely a change of representation. It
produces additional terms in the Hermitian frame and can therefore act as a genuine
driving mechanism. In the present model the scaling part of the Dyson map modulates the
effective coupling, whereas the squeezing part produces a moving-boundary contribution.
Both mechanisms feed into the instantaneous RPA frequencies and can generate
nonadiabatic transitions between oscillator states.

This provides the main difference from our previous work. There the moving-boundary
term opened transitions between spin-boson sectors whose boson numbers differed by two.
Here, after the collective RPA reduction, the induced transitions occur between
instantaneous oscillator states of the RPA branches. The relevant transition amplitude is
controlled by the time-dependence of the RPA frequency $W_j(t)$. In particular, the
leading instantaneous-basis transition $n\to n+2$ is proportional to $\dot W_j/W_j$ and
therefore vanishes in the time-independent case. Thus the transition is not caused by the
static Sch\"utte-Da~Provid\^encia interaction itself, but by the nonadiabatic driving induced
through the time-dependent Dyson map.

The paper is organised as follows. In section 2 we introduce the time-dependent
non-Hermitian Sch\"utte-Da~Provid\^encia Hamiltonian, construct its Hermitian
Dyson-related counterpart and perform the RPA reduction using the
Holstein-Primakoff representation. In section 3 we derive the instantaneous RPA
frequency equation and identify the stable regions in which the collective branches have
real positive frequencies. In section 4 we compute transition probabilities between
instantaneous RPA oscillator states, first in instantaneous-basis perturbation theory and
then exactly within the RPA approximation using Lewis-Riesenfeld invariants. In section 5
we illustrate the two driving mechanisms numerically: modulation of the Dyson-map
scaling parameter and modulation of the squeezing, or moving-boundary, parameter. We
summarise our conclusions in section 6.

\section{Non-Hermitian Sch\"utte-Da~Provid\^encia model in RPA reduction}

The Sch\"utte-Da~Provid\^encia model \cite{schutte1977solvable} describes a bosonic mode coupled to collective particle-hole excitations between two degenerate fermionic shells. In this work we consider a time-dependent non-Hermitian generalisation of the model, defined by the Hamiltonian
\begin{equation}
	H(t)	=	\omega_f(t) S_0	+	\omega_b(t) N_b	+	\alpha(t) S_+ b^\dagger	+	\beta(t) S_- b,
	\qquad	\omega_f(t),\omega_b(t),\alpha(t),\beta(t)\in\mathbb{C}.
	\label{SchdP}
\end{equation}
Throughout we use units in which $\hbar=1$. Here $b^\dagger$ and $b$ are bosonic creation and annihilation operators satisfying
$[b,b^\dagger]=1$, and $N_b=b^\dagger b$ is the boson number operator. The fermionic part of the model consists of two shells, labelled by 1 and 2, each with degeneracy $2\Omega$. We assume that the physical system contains $2\Omega$ fermions. The corresponding collective particle-hole degrees of freedom are described by the collective quasi-spin operators
\begin{align}
	S_0
	&=	\frac{1}{2}	\sum_{k=1}^{2\Omega}	\left(	a_{2k}^\dagger a_{2k}	-	a_{1k}^\dagger a_{1k}	\right)	=	\frac{1}{2}	\left(	N_f-2\Omega	\right),	\	S_+
	&=	\sum_{k=1}^{2\Omega}	a_{2k}^\dagger a_{1k},
	\qquad
	S_- = S_+^\dagger .
\end{align}
The operators $a_{ik}^\dagger$ and $a_{ik}$, with $i=1,2$ and
$k=1,\ldots,2\Omega$, are fermionic creation and annihilation operators.
The quasi-spin generators satisfy the $\mathfrak{su}(2)$-algebra $[S_0,S_\pm]=\pm S_\pm$, $[S_+,S_-]=2S_0$. In the expression for $S_0$, $N_f$ denotes the number of upper-level particles plus lower-level holes, namely
\begin{equation}
	N_f	=	\sum_{k=1}^{2\Omega}	\left(	a_{2k}^\dagger a_{2k}	+	a_{1k}a_{1k}^\dagger	\right).
\end{equation}

The Hermitian Sch\"utte-Da~Provid\^encia model  is recovered when
$\omega_f(t)$ and $\omega_b(t)$ are real and the two interaction strengths satisfy
$\beta(t)=\alpha^*(t)$. In the non-Hermitian case considered here, the parameters are allowed to be complex and, in particular, $\alpha(t)$ and $\beta(t)$ are independent. The latter choice describes an asymmetric coupling between the creation and annihilation of a boson and a collective particle-hole excitation.

In \cite{fring2026induced} we constructed a Hermitian counterpart 
\begin{eqnarray}
	h(t)&=& \left(   \omega_f + i \dot{\delta} \right)S_0 
	+ \left(   \omega_b +  i \dot{\gamma}    \right)
	\left[  \cosh(2 \kappa) N_b  - \frac{1}{2}\sinh(2\kappa) ( b^{\dagger 2} +b^2)  + \sinh^2(\kappa)  \right]  \label{hermh}   \\
	&& \!\!\! + \alpha e^{\gamma + \delta}S_+\left(  b^\dagger  \cosh \kappa  - b \sinh \kappa \right)  
	+ \beta  e^{-(\gamma + \delta )} S_- \left(  b  \cosh \kappa  - b^\dagger \sinh \kappa \right) + i  \frac{\dot{\kappa}}{2} ( b^{\dagger 2} - b^2) ,  \notag
\end{eqnarray}
by solving the  time-dependent Dyson equation
\begin{equation}
	h(t)= \eta(t) H(t) \eta^{-1}(t) + i \dot{\eta}(t) \eta^{-1}(t),   \label{tDyson}
\end{equation}
for the  time-dependent Dyson map
\begin{equation}
\eta(t) := e^{\kappa(t)( b^{\dagger 2} -b^2)/2} e^{\gamma(t) N_b } e^{\delta(t) S_0 } , \qquad \kappa(t),\gamma(t), \delta(t) \in \mathbb{C} , \label{Dysonm}
\end{equation}
Here and below the time dependence of \(\gamma,\delta,\kappa\) is understood.

The Hamiltonian $h(t)$ is Hermitian provided the conditions
\begin{equation}
	A_f(t):=\omega_f + i \dot{\delta} \in \mathbb{R}, \qquad  
	A_b(t):=\omega_b +  i \dot{\gamma}  \in \mathbb{R}, \qquad
	\kappa  \in \mathbb{R}, \qquad 
	\beta = \alpha^*  e^{2 \Re (\gamma + \delta )} . \label{hermcond}
\end{equation}
In addition, the metric and its inverse remain bounded, for instance, when
\(\gamma(t)\in i\mathbb{R}\). These conditions ensure that the transformed
Hamiltonian defines a unitary evolution in the Dyson-related Hermitian frame.

For our purposes here it is useful to introduce the squeezed bosonic operators by means of a Bogoliubov transformation
\begin{equation}
	B^\dagger
	=
	b^\dagger\cosh\kappa(t)
	-
	b\sinh\kappa(t),
	\qquad
	B
	=
	b\cosh\kappa(t)
	-
	b^\dagger\sinh\kappa(t), 
\end{equation}
satisfying the usual bosonic commutation relation $	[B,B^\dagger]=1 $.
In terms of these operators the Hamiltonian takes the compact form
\begin{equation}
	h(t)
	=
	A_f(t)S_0
	+
	A_b(t)B^\dagger B
	+
	g(t)S_+B^\dagger
	+
	g^*(t)S_-B
	+
	\frac{i\dot\kappa(t)}{2}
	\left(
	B^{\dagger 2}-B^2
	\right) ,
	\label{Compact}
\end{equation}
where $g(t):= \alpha(t) e^{\gamma(t) + \delta(t)}$. This form makes the role of the Dyson map transparent. The function
\(\delta(t)\) rescales the effective coupling \(g(t)\), while \(\kappa(t)\)
enters both through the squeezed boson \(B\) and through the explicitly
time-dependent boundary term proportional to \(\dot\kappa(t)\).

We now pass to a random phase approximation \cite{bohm1951collective,pines1952collective,bohm1953collective}, see e.g. \cite{co2023introducing} for a recent review. The basic assumption is that the
collective sector dominates the dynamics. To leading order, the fermionic
operators may be bosonised by means of the Holstein-Primakoff representation \cite{Holstein},
\begin{align}
	S_+
	&=
	\sqrt{2\Omega}\,
	a^\dagger
	\left(
	1-\frac{a^\dagger a}{2\Omega}
	\right)^{1/2}
	\simeq
	\sqrt{2\Omega}\,a^\dagger ,   \label{RPAa1}
	\\
	S_-
	&=
	\sqrt{2\Omega}
	\left(
	1-\frac{a^\dagger a}{2\Omega}
	\right)^{1/2}
	a
	\simeq
	\sqrt{2\Omega}\,a ,
	\\
	S_0
	&=
	a^\dagger a-\Omega   \label{RPAa3} .
\end{align}
This RPA approximation is valid for low collective occupations
\begin{equation}
	\langle a^\dagger a\rangle \ll 2\Omega .
\end{equation}
In the commutators generated by the interaction terms we use, consistently with
$a^\dagger a\ll 2\Omega$, the leading approximation
$S_0\simeq -\Omega$.
Up to an irrelevant constant, the Hamiltonian \eqref{Compact} then becomes
a quadratic bosonic Hamiltonian. The corresponding collective RPA excitation is
taken in the form
\begin{equation}
	\Gamma^\dagger
	:=
	X_1S_+
	-
	Y_1S_-
	+
	X_2B^\dagger
	-
	Y_2B ,
	\label{RPAphonon}
\end{equation}
with normalisation condition
\begin{equation}
	[\Gamma,\Gamma^\dagger]=1 ,
\end{equation}
which holds when the constants $X_1,X_2,Y_1,Y_2$  are constrained as
\begin{equation}
	2\Omega
	\left(
	|X_1|^2-|Y_1|^2
	\right)
	+
	|X_2|^2-|Y_2|^2
	=
	1 .
\end{equation}
The RPA frequencies $W(t)$ are determined by the equation of motion
\begin{equation}
	[h(t),\Gamma^\dagger]
	=
	W(t)\Gamma^\dagger .
	\label{RPAeom}
\end{equation}
This is the instantaneous RPA eigenvalue problem for the collective mode. It
leads to the secular equation for the two RPA frequencies \(W_\pm(t)\), which
will be analysed in the next section. Only those branches for which the
frequencies are real and positive give stable oscillator modes.

For each stable RPA branch the collective Hamiltonian may be represented as a
time-dependent harmonic oscillator. Introducing canonical variables
\(\hat x_j,\hat p_j\), one may write
\begin{equation}
	\Gamma_j^\dagger	=	\sqrt{\frac{mW_j(t)}{2}}	\left(	\hat x_j	-	\frac{i}{mW_j(t)}\hat p_j	\right),
	\qquad
	\Gamma_j	=	\sqrt{\frac{mW_j(t)}{2}}	\left(	\hat x_j	+	\frac{i}{mW_j(t)}\hat p_j	\right),
\end{equation}
so that the RPA Hamiltonian for branch \(j=\pm\) takes the form of the harmonic oscillator with time-dependent frequency
\begin{equation}
	H_{\mathrm{RPA},j}(t)
	=
	\frac{\hat p_j^{\,2}}{2m}
	+
	\frac{m}{2}W_j^2(t)\hat x_j^{\,2}.
	\label{HRPA}
\end{equation}
Thus the time-dependent non-Hermitian spin-boson problem, after the Dyson
transformation and the RPA reduction, is mapped to two time-dependent oscillator
branches. This oscillator form can be solved exactly using Lewis-Riesenfeld invariants \cite{Lewis69,Pedrosa,AndTom2}. It is the starting point for both the perturbative and the exact transition-probability calculations below.

\section{Stable RPA regions}

The equation of motion \eqref{RPAeom}  can be written explicitly as a finite-dimensional
RPA eigenvalue problem. Suppressing the time dependence and using the RPA
replacements (\ref{RPAa1})-(\ref{RPAa3}), the relevant commutators are
\begin{align}
	[h,S_{+}]_{\mathrm{RPA}}
	&=
	A_f S_{+} +2\Omega g^{*} B,
	&
	[h,S_{-}]_{\mathrm{RPA}}
	&=
	-A_f S_{-} -2\Omega g B^{\dagger},
	\\
	[h,B^{\dagger}]
	&=
	A_b B^{\dagger} + g^{*}S_{-} - i\dot{\kappa} B,
	&
	[h,B]
	&=
	-A_b B - gS_{+} - i\dot{\kappa} B^{\dagger}.
\end{align}
Substitution of $\Gamma^{\dagger}$ from (\ref{RPAphonon}) then gives
\begin{equation}
	\mathcal{M}_{\mathrm{RPA}}
	\begin{pmatrix}
		X_1\\
		Y_1\\
		X_2\\
		Y_2
	\end{pmatrix}
	=
	W
	\begin{pmatrix}
		X_1\\
		Y_1\\
		X_2\\
		Y_2
	\end{pmatrix},
\end{equation}
where
\begin{equation}
	\mathcal{M}_{\mathrm{RPA}}
	=
	\begin{pmatrix}
		A_f & 0 & 0 & g\\
		0 & -A_f & -g^{*} & 0\\
		0 & 2\Omega g & A_b & i\dot{\kappa}\\
		-2\Omega g^{*} & 0 & i\dot{\kappa} & -A_b
	\end{pmatrix}.
\end{equation}
The secular condition $\det\left({\cal M}_{\rm RPA}-W \mathbb{I}\right)=0$ then yields the determining equation for the RPA frequencies
\begin{equation}
\left(W^2-A_f^2\right)
\left(\dot\kappa^2+W^2-A_b^2\right)
+
4\Omega |g|^2\left(W^2-A_bA_f\right)
+
4\Omega^2|g|^4
=0 .
\end{equation}
Solving this fourth order equation and keeping only the positive frequencies gives
 \begin{equation}
	W_\pm(t)
	=
	\sqrt{
		\frac{
			A_f(t)^2+A_b(t)^2-\dot\kappa(t)^2-4\Omega |g(t)|^2
			\pm
			\sqrt{\Delta(t)}
		}{2}
	}
 \end{equation}
with discriminant
 \begin{equation}
	\Delta(t)
	=
	\left[
	A_f(t)^2-A_b(t)^2+\dot\kappa(t)^2
	\right]^2
	-
	8\Omega |g(t)|^2
	\left[
	\left(A_f(t)-A_b(t)\right)^2-\dot\kappa(t)^2
	\right].
 \end{equation}
Only those branches for which $\Delta(t)\geq 0$, $W_\pm(t)^2>0$ are stable real oscillator frequencies.
A rough estimate of the allowed region is obtained from the special case $\dot\kappa(t)=0$, for which the discriminant becomes
 \begin{equation}
\Delta_0
=
(A_f-A_b)^2
\left[
(A_f+A_b)^2-8\Omega |g|^2
\right].
 \end{equation}
Thus, the static estimate is
 \begin{equation}
	8\Omega |g|^2\leq (A_f+A_b)^2 .
 \end{equation}
At equality, one of the RPA modes may become soft. Beyond this region, the RPA frequencies become complex, signalling an instability or breakdown of the RPA approximation. For genuinely time-dependent boundaries this condition should only be used as a
static estimate. In the numerical analysis below we impose the full instantaneous
conditions $\Delta(t)\geq0$ and $W_\pm^2(t)>0$ over the full time interval.

\section{RPA transition probabilities: perturbative and exact}

\subsection{Instantaneous-basis perturbation theory}

At each fixed time $t$, the instantaneous eigenvalue problem for the RPA Hamiltonian $H_{\mathrm{RPA},j}(t)$ in (\ref{HRPA}) is
 \begin{equation}
H_{\mathrm{RPA},j}(t)|n,j;t\rangle
=
E_{n,j}(t)|n,j;t\rangle , \label{insteig}
 \end{equation}
with instantaneous eigenstates $|n,j;t\rangle$ and instantaneous eigenenergies
 \begin{equation}
E_{n,j}(t) =  W_j(t)\left(n+\frac12\right), \qquad n=0,1,2,\ldots .  \label{insteigv}
 \end{equation}
We expand an arbitrary state in the instantaneous basis as
 \begin{equation}
|\Psi_j(t)\rangle =\sum_n c_{n}^{(j)}(t) \exp\!\left[ -i \int_0^t ds\,E_{n,j}(s) \right] |n,j;t\rangle .
 \end{equation}
Substituting this into the Schrödinger equation
 \begin{equation}
i\frac{d}{dt}|\Psi_j(t)\rangle = H_{\mathrm{RPA},j}(t)|\Psi_j(t)\rangle
 \end{equation}
gives
 \begin{equation}
\dot c_m^{(j)}(t)=-\sum_n c_n^{(j)}(t) \langle m,j;t|\partial_t n,j;t\rangle \exp\!\left[ i \int_0^t ds\,
\bigl(E_{m,j}(s)-E_{n,j}(s)\bigr)  \right].
 \end{equation}
To first order, for an initial state $|n,j;0\rangle$, we set
 \begin{equation}
c_n^{(j)}(t)\simeq 1,
\qquad
c_{m\neq n}^{(j)}(t)\simeq 0.     \label{firstord}
 \end{equation}
Hence the transition amplitude from $n$ to $m$ is
 \begin{equation}
A^{(j)}_{m\leftarrow n}(T)
\simeq
-\int_0^T dt\,
\langle m,j;t|\partial_t n,j;t\rangle
\exp\!\left[
i \int_0^t ds\,
\bigl(E_{m,j}(s)-E_{n,j}(s)\bigr)
\right].
 \end{equation}
 Next we compute the nonadiabatic matrix element $\langle m,j;t|\partial_t n,j;t\rangle$. We start by differentiating the instantaneous eigenvalue equation (\ref{insteig}). Projecting with $\langle m,j;t|$ for $m\neq n$, gives
 \begin{equation}
 \langle m,j;t|\partial_t n,j;t\rangle
 =
 \frac{\langle m,j;t|\partial_t H_{\mathrm{RPA},j}(t)|n,j;t\rangle}
 {E_{n,j}(t)-E_{m,j}(t)} .
 \end{equation}
With the expression (\ref{HRPA}) for the RPA Hamiltonian we get
 \begin{equation}
 \partial_t H_{\mathrm{RPA},j}(t)
 =
 mW_j(t)\dot W_j(t)\hat x_j^2 .
 \end{equation}
 Using the instantaneous oscillator representation
 \begin{equation}
 \hat x_j
 =
 \sqrt{\frac{1}{2mW_j(t)}}
 \left(a_j+a_j^\dagger\right),
 \end{equation}
we obtain
 \begin{equation}
 \langle n+2,j;t|\hat x_j^2|n,j;t\rangle
 =
 \frac{1}{2mW_j(t)}
 \sqrt{(n+1)(n+2)} .
 \end{equation}
 Therefore
 \begin{equation}
 \langle n+2,j;t|\partial_t H_{\mathrm{RPA},j}(t)|n,j;t\rangle
 =  \frac{ \dot W_j(t)}{2}  \sqrt{(n+1)(n+2)} .
 \end{equation}
 Together with
 \begin{equation}
 	E_{n+2,j}(t)-E_{n,j}(t)
 	=
 	2 W_j(t).
 \end{equation}
 this gives, up to a phase convention, for the instantaneous eigenstates,
 \begin{equation}
 \langle n+2,j;t|\partial_t n,j;t\rangle
 =
 -\frac{\dot W_j(t)}{4W_j(t)}
 \sqrt{(n+1)(n+2)} .   \label{ntonp2}
 \end{equation}
 With the opposite phase convention this matrix element appears with the opposite sign.
 The sign is immaterial for the transition probability. Equation (\ref{ntonp2}) shows explicitly
 that the $n\to n+2$ transition is purely nonadiabatic: it is generated by the time
 dependence of the instantaneous RPA basis and is proportional to $\dot W_j(t)$. Hence,
 in the time-independent case, where $W_j$ is constant, no $n\to n+2$ transition is
 induced.

Thus we have
 \begin{equation}
A^{(j)}_{n+2\leftarrow n}(T)
\simeq
-\frac{\sqrt{(n+1)(n+2)}}{4}
\int_0^T dt\,
\frac{\dot W_j(t)}{W_j(t)}
\exp\!\left[
2i\int_0^t ds\,W_j(s)
\right].
 \end{equation}

The corresponding transition probability is
 \begin{equation}
P^{(j)}_{n+2\leftarrow n}(T)
\simeq
\left|
A^{(j)}_{n+2\leftarrow n}(T)
\right|^2 .    \label{transpert}
 \end{equation}

This expression should be understood as a first-order perturbative transition probability. It was obtained by expanding the state in the instantaneous oscillator basis and retaining only the leading nonadiabatic coupling between $|n,j;t\rangle$ and $|n+2,j;t\rangle$. Consequently, the result is reliable only when the corresponding transition amplitude remains small, i.e. $ \left|A^{(j)}_{n+2\leftarrow n}(T)\right|\ll 1 $. Thus if the expression for $P^{(j)}_{n+2\leftarrow n}(T)$ becomes comparable to or larger than unity, this signals the breakdown of the first-order approximation rather than a physical probability exceeding one or a missing normalisation factor. In addition, in the RPA setting we need to remain within the range of validity of the bosonised approximation, namely occupations satisfying $n\ll 2\Omega$.

\subsection{Exact Lewis-Riesenfeld transition amplitudes}

Using the RPA oscillator reduction described above, we can compare the perturbative analysis with the exact analytic solution within the RPA approximation. For a given branch \(j\), the exact solution that solves the full time-dependent Schr\"odinger equation can be written in terms of the Lewis-Riesenfeld invariant 
\begin{equation}
	\Psi_n^{(j)}(x,t)
	=
	e^{-i(n+1/2)\theta_j(t)}
	\frac{1}{\sqrt{2^n n!}}
	\left(
	\frac{1}{\pi\rho_j^2(t)}
	\right)^{1/4}
	H_n\!\left(
	\frac{x}{\rho_j(t)}
	\right)
	e^{  \frac{ 1 }{2 \rho_j^2(t)}  \left[
	i m\dot \rho_j(t)  \rho_j (t) -1 \right] x^2},
\end{equation}
where
\begin{equation}
	\theta_j(t)
	=
	\int_0^t\frac{dt'}{m\rho_j^2(t')} ,
\end{equation}
and  \(\rho_j(t)\) is a real solution of the
Ermakov-Pinney equation
\begin{equation}
	\ddot\rho_j(t)+W_j^2(t)\rho_j(t)
	=
	\frac{1}{m^2\rho_j^3(t)} .
\end{equation}
The eigenfunctions of the instantaneous Hamiltonian $H_j(t)$ are the ordinary harmonic-oscillator eigenfunctions with frequency $W_j(t)$.
\begin{equation}
	\phi_{n,j}(x;t)
	=
	\frac{1}{\sqrt{2^n n!}}
	\left(
	\frac{mW_j(t)}{\pi}
	\right)^{1/4}
	H_n\!\left(
	\sqrt{  mW_j(t) }\,x
	\right)
	\exp\!\left[
	-\frac{mW_j(t)x^2}{2}
	\right].
\end{equation}
Thus, choosing the initial values 
\begin{equation}
	\rho_j(0)=\frac{1}{\sqrt{mW_j(0)}},
	\qquad
	\dot\rho_j(0)=0 ,
\end{equation}
ensures  that $\Psi_n^{(j)}(x,0) =	\phi_{n,j}(x;0) $. Then, the exact transition amplitude from the initial instantaneous state $\phi_{n,j}(x;0)$ to the state $\phi_{n+2,j}(x;T)$ is given as
\begin{equation}
	A_{n+2\leftarrow n}^{(j)}(T)
	=
	\int_{-\infty}^{\infty} dx\,
	\phi_{n+2,j}^{*}(x;T)\,
	\Psi_n^{(j)}(x,T).
\end{equation}
Next we compute this expression. Suppressing the branch label \(j\) and abbreviating at the
final time \(T\)
\begin{equation}
	q:=\sqrt{mW(T)},
	\quad
	r:=\frac{1}{\rho(T)},
	\quad
	s:=m\frac{\dot\rho(T)}{\rho(T)}, \quad 	C:=q^2+r^2-is .
\end{equation}
we obtain
\begin{equation}
	A_{n+2\leftarrow n}(T)
	=
	e^{-i(n+1/2)\theta(T)}
	\frac{1}{\sqrt{2^{2n+2}(n+2)!n!}}
	\sqrt{\frac{qr}{\pi}}\,
	I_n ,
\end{equation}
where
\begin{equation}
	I_n
	=
	\int_{-\infty}^{\infty} dx\,
	H_{n+2}\!\left(   qx    \right)
	H_n\!\left(    rx   \right)
	\exp\!\left[
	-\frac{C x^2}{2}
	\right].
\end{equation}
We evaluate this integral using the generating function for the Hermite polynomials
\begin{equation}
e^{-u^2+2uy}= \sum_{m=0}^{\infty}\frac{u^m}{m!}H_m(y).
\end{equation}
Accordingly, we can express the two Hermite polynomials in the integrand as
\begin{equation}
H_{n+2}\left(   qx      \right)
= \left.\partial_u^{\,n+2} e^{-u^2+2uqx } \right|_{u=0}, \qquad
H_n\left(  rx  \right)
= \left.\partial_v^{\,n} e^{-v^2+2vrx } \right|_{v=0}.
\end{equation}
Therefore, the integral becomes
\begin{equation}
I_n
=
\left.
\partial_u^{\,n+2}\partial_v^{\,n}
\int_{-\infty}^{\infty}
dx\,
\exp\left[
-\frac{Cx^2}{2}
-u^2-v^2
+  2x (qu+rv)
\right]
\right|_{u=v=0}.
\end{equation}
The \(x\)-integral is elementary, so that we obtain
 \begin{equation}
I_n = \sqrt{\frac{2\pi}{C}} \left. \frac{\partial^{n+2}}{\partial u^{n+2}} \frac{\partial^n}{\partial v^n} \exp\!\left[ a u^2+cuv+bv^2 \right] \right|_{u=v=0},
\end{equation} 
with
 \begin{equation} 
 	a=\frac{2q^2}{C} -1, \qquad b=\frac{2r^2}{C}-1, \qquad c=\frac{4qr}{C}. 
 \end{equation}
  Expanding the exponential functions as
  \begin{equation} 
  	e^{a u^2} = \sum_{p=0}^{\infty}\frac{a^p u^{2p}}{p!}, \qquad 
  	e^{cuv} = \sum_{k=0}^{\infty}\frac{c^k u^k v^k}{k!}, \qquad 
  	e^{bv^2} = \sum_{\ell=0}^{\infty}\frac{b^\ell v^{2\ell}}{\ell!}, 
  	\end{equation} 
  	we note that the derivative \(\partial_u^{n+2}\partial_v^n\) selects the coefficient of \(u^{n+2}v^n\), so that we require 
	\begin{equation} 
  		2p+k=n+2, \quad k+2\ell=n, \quad \Rightarrow \quad p=\ell +1, \quad \ell=0,1,\ldots,\left\lfloor \frac{n}{2}\right\rfloor .
  	\end{equation} 
  	The coefficient of \(u^{n+2}v^n\) is therefore 
 \begin{equation} 
  		\sum_{\ell=0}^{\lfloor n/2\rfloor} \frac{ a^{\ell+1}b^\ell c^{\,n-2\ell} }{ (\ell+1)!\,\ell!\,(n-2\ell)! },
 \end{equation} 
  	so that
  	\begin{equation}
  		 I_n = \sqrt{\frac{2\pi}{C}}\, (n+2)!n! \sum_{\ell=0}^{\lfloor n/2\rfloor} \frac{ a^{\ell+1}b^\ell c^{\,n-2\ell} }{ (\ell+1)!\,\ell!\,(n-2\ell)! }. 
  		 \end{equation} 
  		Restoring the branch label \(j\), the exact RPA transition amplitude becomes 
  		\begin{equation} 
  		A_{n+2\leftarrow n}^{(j),{\rm exact}}(T) = e^{-i(n+1/2)\theta_j(T)} \sqrt{\frac{2q_jr_j}{C_j}}\, \frac{\sqrt{(n+2)!n!}}{2^{n+1}} \sum_{\ell=0}^{\lfloor n/2\rfloor} \frac{ a_j^{\ell+1}b_j^\ell c_j^{\,n-2\ell} }{ (\ell+1)!\,\ell!\,(n-2\ell)! } .
  		\end{equation}
  			For the special cases \(n=0\) and $n=2$ the expressions simplify to 
  	\begin{eqnarray} 
  			A_{2\leftarrow 0}^{(j),{\rm exact}}(T) &=& e^{-i\theta_j(T)/2} \sqrt{\frac{q_jr_j}{C_j}}\, a_j , \label{02exact}  \\
  			A_{4\leftarrow 2}^{(j),{\rm exact}}(T) &=&  \frac{1}{2} \sqrt{\frac{3}{2}}   e^{-5 i\theta_j(T)/2} \sqrt{\frac{q_jr_j}{C_j}}\,  a_j  (a_j  b_j + c_j^2  )  .
  		\end{eqnarray} 
The corresponding exact RPA transition probability is therefore 
\begin{equation} 
	P_{n+2\leftarrow n}^{(j),{\rm exact}}(T) = \left| A_{n+2\leftarrow n}^{(j),{\rm exact}}(T) \right|^2 .  \label{transexact}
\end{equation} 
\subsection{Weak-driving limit and comparison}

In the weak-driving regime the Lewis-Riesenfeld result reduces to the first-order instantaneous-basis expression. The reason is that a slowly or weakly varying frequency $W_j(t)$ produces only a small mismatch between the exact evolved state and the final instantaneous oscillator basis. Equivalently, the squeezing parameters encoded in $a_j,b_j,c_j$ remain small, and the dominant contribution to the transition amplitude is the direct $n\to n+2$ nonadiabatic coupling proportional to $\dot W_j/W_j$. Thus the perturbative expression captures the leading resonance structure, while the exact expression contains the higher-order squeezing corrections and the depletion of the initial state. This distinction becomes important near strong resonances, where the perturbative probability can overestimate the exact transition probability even though it continues to predict the correct peak positions.

\section{Numerical analysis}
We now use the expressions from the previous section to analyse two complementary driving mechanisms. 
The first mechanism is driven by the scaling part of the Dyson map: the parameter
$\delta(t)$ is modulated in time and its modulation frequency is scanned, while the
boundary protocol $\kappa(t)$ is kept fixed. Since $\delta(t)$ enters the effective coupling $|g(t)|$, this produces a time-dependent RPA frequency and hence nonadiabatic transitions between 
instantaneous oscillator states. The second mechanism is driven by the 
squeezing part of the Dyson map: $\delta$ is kept fixed, but the boundary 
parameter $\kappa(t)$ is varied periodically. In this case the transition is 
generated directly by the moving-boundary contribution to $W_+(t)$. In both 
cases the peaks in the transition probability can be understood as parametric 
resonances, while the smaller neighbouring peaks arise from sideband mixing 
with the additional modulation of $A_b(t)$.

\subsection{Dyson-map driven transition probability}

\begin{figure}[h]
	\begin{minipage}[b]{\textwidth}      
		\centering
		\includegraphics[width=0.87\textwidth]{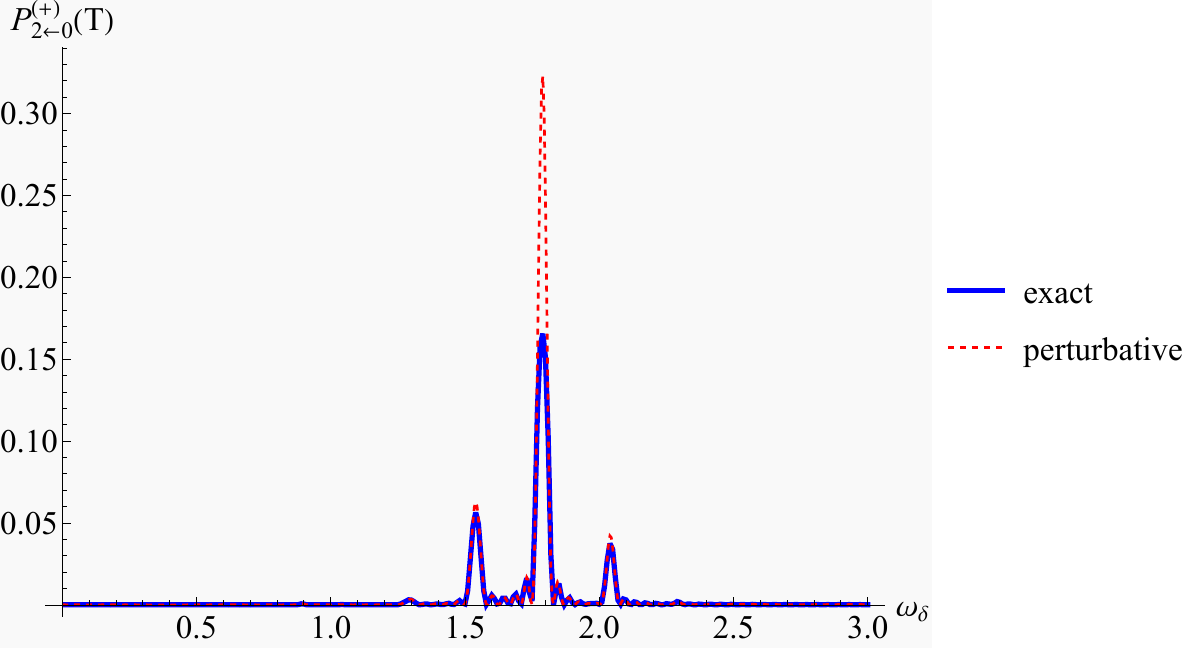}
	\end{minipage}   
	\caption{
		Dyson-map driven transition probability $P^{(+)}_{2\leftarrow0}(T)$ in the perturbative approximation (\ref{transpert}) and exact analytical form  (\ref{transexact}) as a
		function of the modulation frequency $\omega_\delta$. The parameters are
		$A_f(t)=0$, $A_b(t)=1+0.08\cos(\omega_b t)$,
		$\kappa(t)=\kappa_0\sin(\omega_\kappa t)$,
		$\delta(t)=\delta_0+d_0\sin(\omega_\delta t)$,
		$|g(t)|=\lambda e^{\delta(t)}/\sqrt{2\Omega}$,
		$\omega_b=0.25$, $\kappa_0=0.05$, $\omega_\kappa=0.8$,
		$\delta_0=0.2$, $d_0=0.05$, $\lambda=0.25$,
		$T=2\pi 20/\omega_\kappa$, and $\Omega=25$.
		The dominant peak occurs near $\omega_\delta\simeq1.79$, while the smaller
		sideband peaks occur near $\omega_\delta\simeq1.54$ and
		$\omega_\delta\simeq2.04$.
	}
	\label{deltatrans}
\end{figure}

Figure \ref{deltatrans} displays the RPA transition probability
$P^{(+)}_{2\leftarrow0}(T)$ as a function of the Dyson-map modulation
frequency $\omega_\delta$ in the perturbative approximation (\ref{transpert}) and exact analytical form  (\ref{transexact}). The parameter $\delta(t)=\delta_0+d_0\sin(\omega_\delta t)$
enters the effective coupling as
\begin{equation}
	|g(t)|=\frac{\lambda e^{\delta(t)}}{\sqrt{2\Omega}},
\end{equation}
and therefore modulates the instantaneous RPA frequency $W_+(t)$. This time
dependence generates the nonadiabatic coupling between instantaneous oscillator
states responsible for the transition $0\to2$.

The dominant peak occurs when the modulation frequency satisfies the parametric
resonance condition
\begin{equation}
	\omega_\delta \simeq 2\overline W_+,
	\qquad
	\overline W_+=\frac1T\int_0^T W_+(t)\,dt .
\end{equation}
This reflects the fact that the transition $0\to2$ has energy gap
$2 W_+$. The smaller peaks arise as sideband resonances, generated by
the additional modulation of the RPA frequency through $A_b(t)=1+0.08\cos(\omega_b t)$.
They are expected near
\begin{equation}
	\omega_\delta\simeq 2\overline W_+\pm\omega_b .
\end{equation}
For the parameters of figure \ref{deltatrans}, this gives sideband peaks near
$\omega_\delta\simeq1.54$ and $\omega_\delta\simeq2.04$, surrounding the
dominant resonance near $\omega_\delta\simeq1.79$. The figure therefore shows
that the transition can be driven directly by the time-dependent Dyson map. By
tuning $\omega_\delta$, the integrated transition amplitude can be enhanced
or suppressed through constructive or destructive phase interference.

We find that the exact Lewis-Riesenfeld result agrees well with the first-order nonadiabatic expression in the weak-driving regime. For stronger driving, especially near the dominant resonance, the perturbative result overestimates the transition probability as seen in figure \ref{deltatrans}  because the first-order approximation neglects depletion of the initial state, i.e. it keeps $c_n^{(j)}(t)\simeq 1$ in (\ref{firstord}), and higher-order squeezing-induced redistribution among the even oscillator levels. This explains why the perturbative and exact curves have the same resonance positions but different peak heights.

\subsection{Boundary driven transition probability}

\begin{figure}[h]
	\begin{minipage}[b]{\textwidth}      
		\centering
		\includegraphics[width=0.87\textwidth]{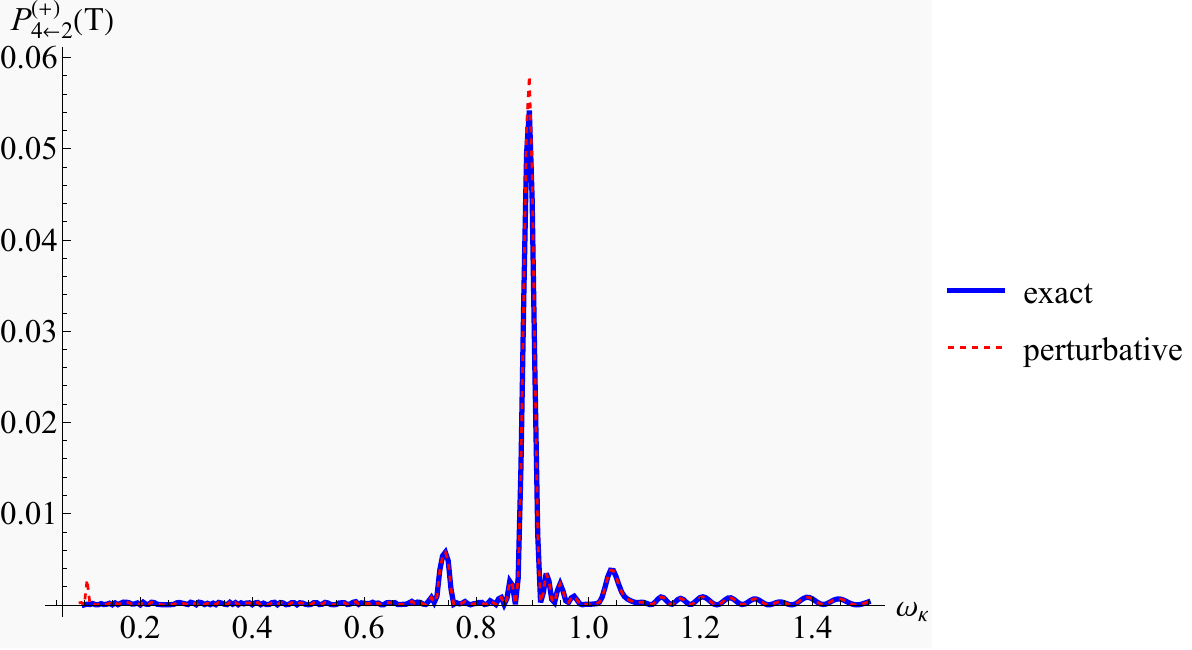}
	\end{minipage}   
	\caption{
		Boundary driven transition probability $P^{(+)}_{4\leftarrow2}(T)$ in the perturbative approximation (\ref{transpert}) and exact analytical form  (\ref{transexact}) as a
		function of the modulation frequency $\omega_\kappa$. The parameters are
		$A_f(t)=0$, $A_b(t)=1+0.08\cos(\omega_b t)$,
		$\kappa(t)=\kappa_0\sin(\omega_\kappa t)$,
		$\delta(t)=\delta_0$,
		$|g(t)|=\lambda e^{\delta(t)}/\sqrt{2\Omega}$,
		$\omega_b=0.3$, $\kappa_0=0.1$, $\delta_0=0.2$,
		$\lambda=0.25$, $T=2\pi 20/\omega_\kappa$, and $\Omega=20$.
		The dominant peak occurs near $\omega_\kappa\simeq0.89$, while the smaller
		sideband peaks occur near $\omega_\kappa\simeq0.74$ and
		$\omega_\kappa\simeq1.04$.
	}
	\label{boundarytrans}
\end{figure}

Figure~\ref{boundarytrans} displays the RPA transition probability
$P^{(+)}_{4\leftarrow2}(T)$ as a function of the boundary modulation frequency
$\omega_\kappa$ in the perturbative approximation (\ref{transpert}) and exact analytical form  (\ref{transexact}). In this case the Dyson map scaling parameter is kept fixed, $\delta(t)=\delta_0$,
so that the effective coupling
\begin{equation}
	|g(t)|=\frac{\lambda e^{\delta_0}}{\sqrt{2\Omega}}
\end{equation}
is time independent. The transition is instead driven by the squeezing part of the Dyson map through $
\kappa(t)=\kappa_0\sin(\omega_\kappa t)$. Since the instantaneous RPA frequency depends on $\dot\kappa(t)^2$, the boundary motion contributes through
\begin{equation}
	\dot\kappa(t)^2
	=
	\kappa_0^2\omega_\kappa^2\cos^2(\omega_\kappa t)
	=
	\frac{\kappa_0^2\omega_\kappa^2}{2}
	\left[1+\cos(2\omega_\kappa t)\right].
\end{equation}
Thus the moving boundary modulates $W_+(t)$ predominantly at frequency $2\omega_\kappa$, generating the nonadiabatic coupling between instantaneous oscillator states responsible for the transition $2\to4$.

The dominant peak occurs when the boundary modulation satisfies the parametric resonance condition
\begin{equation}
	2\omega_\kappa\simeq 2\overline W_+,
	\qquad
	\overline W_+=\frac1T\int_0^T W_+(t)\,dt , 
\end{equation}
or equivalently $ \omega_\kappa\simeq \overline W_+ $.
This reflects the fact that the transition $2\to4$ has energy gap $2 W_+$, while the boundary drive enters through the second harmonic $2\omega_\kappa$. The smaller peaks arise as sideband resonances, generated by the additional modulation of the RPA frequency through $A_b(t)=1+0.08\cos(\omega_b t)$.
They are expected near
\begin{equation}
	2\omega_\kappa\simeq 2\overline W_+\pm\omega_b,
\end{equation}
or equivalently $
\omega_\kappa\simeq \overline W_+\pm \omega_b/2$.
For the parameters of figure \ref{boundarytrans}, this gives sideband peaks near
$\omega_\kappa\simeq0.74$ and $\omega_\kappa\simeq1.04$, in agreement with the numerical plot. The figure therefore shows that the transition can also be driven directly by the moving-boundary part of the Dyson map. By tuning $\omega_\kappa$, the integrated transition amplitude can be enhanced or suppressed through constructive or destructive phase interference.

The comparison between the exact and the perturbative solution is similar as in the previous subsection.

\section{Conclusion}

We have studied a time-dependent non-Hermitian generalisation of the
Sch\"utte-Da~Provid\^encia model and analysed its collective dynamics after a
Dyson transformation and an RPA reduction. The time-dependent Dyson map maps
the original non-Hermitian Hamiltonian to a Hermitian counterpart, but it does
more than provide an equivalent representation. Through its explicit time
dependence it generates effective driving terms in the Hermitian frame. After
the RPA reduction these terms appear as time-dependent oscillator frequencies
for the collective modes.

The resulting RPA eigenvalue problem yields two instantaneous oscillator
branches. Their stability is determined by the reality and positivity of the
corresponding frequencies $W_\pm(t)$. Within the stable regions the collective
problem can therefore be reduced to time-dependent harmonic oscillators. This
identification makes it possible to compute transition probabilities between
instantaneous RPA oscillator states. In the perturbative instantaneous-basis
approach, the leading transition amplitude is proportional to
$\dot W_j(t)/W_j(t)$. Thus the transition $n\to n+2$ is purely
nonadiabatic and disappears in the time-independent case, where the RPA basis is
stationary.

We also derived exact transition amplitudes within the RPA approximation by
using Lewis-Riesenfeld invariants. This provides a non-perturbative benchmark
for the first-order instantaneous-basis result. In the weak-driving regime the
two descriptions agree well and predict the same resonance positions. Near
strong resonances, however, the perturbative expression overestimates the
transition probability because it neglects depletion of the initial state and
higher-order squeezing-induced redistribution among oscillator levels.

The numerical examples illustrate two distinct mechanisms by which the
time-dependent Dyson map can drive collective transitions. A modulation of the
scaling parameter $\delta(t)$ changes the effective coupling $g(t)$ and hence
modulates the RPA frequency directly. By contrast, a modulation of the squeezing
parameter $\kappa(t)$ acts as a moving-boundary contribution and drives the
system through the term proportional to $\dot\kappa^2(t)$. In both cases the
dominant transition peaks are well described as parametric resonances, while
the additional smaller peaks arise from sideband mixing with the modulation of
the bosonic frequency.

The main conclusion is therefore that a time-dependent Dyson map can act as a
controlled driving mechanism for collective RPA modes. Even when the transformed
Hamiltonian is Hermitian and the evolution is unitary in the Dyson-related
frame, the explicit time dependence of the map leaves observable dynamical
signatures in the form of induced collective transitions.

\newif\ifabfull\abfulltrue

\end{document}